\documentclass[%
 reprint,
 amsmath,amssymb,
 aps,
]{revtex4-1}

\usepackage{graphicx}
\usepackage{dcolumn}
\usepackage{bm}

\begin{document}

\preprint{APS/123-QED}

\title{ Parameters optimization and real-time calibration of Measurement-Device-Independent Quantum Key Distribution Network based on Back Propagation Artificial Neural Network}
	\author{Feng-Yu Lu}
	\author{Zhen-Qiang Yin}
	\email{yinzq@ustc.edu.cn}
	\author{Chao-Han Cui}
	\author{Chao Wang}
	\author{Jun Teng}
	\author{Shuang Wang}
	\author{Wei Chen}
	\author{De-Yong He}
	\author{Guang-Can Guo}
	\author{Zheng-Fu Han}
	\affiliation{CAS Key Laboratory of Quantum Information, University of Science and Technology of China, Hefei 230026, P. R. China}
	\affiliation{Synergetic Innovation Center of Quantum Information $\&$ Quantum Physics, University of Science and Technology of China, Hefei, Anhui 230026, P. R. China}
	\affiliation{State Key Laboratory of Cryptology, P. O. Box 5159, Beijing 100878, P. R. China}




\date{\today}

\begin{abstract}
The parameters choosing (such as probabilities of choosing X-basis or Z-basis, intensity of signal state and decoy state, etc.) and system calibrating will be more challenging when the number of users of a measurement-device-independent quantum key distribution(MDI-QKD) network becomes larger. At present, people usually use optimization algorithms to search the best parameters. This method can find the optimized parameters accurately but may cost lots of time and hardware resources. It's a big problem in large scale MDI-QKD network. Here, we present a new method, using Back Propagation Artificial Neural Network(BPNN) to predict, rather than searching the optimized parameters. Compared with optimization algorithms, our BPNN is faster and more lightweight, it can save system resources. Another big problem brought by large scale MDI-QKD network is system recalibration. BPNN can support this work in real time, and it only needs to use some discarded data generated from communication process, rather than require us to add additional devices or scan the system

\end{abstract}

\pacs{Valid PACS appear here}
\maketitle


\section{Introduction}
Quantum key distribution(QKD) allows two remote users, Alice and Bob, share random keys with information-theoretic security \cite{shor2000simple,lo1999unconditional}. However, there are still some loopholes since imperfect of devices\cite{lutkenhaus2000security,brassard2000limitations,fung2007phase,xu2010experimental,zhao2008quantum,gerhardt2011full,makarov2009controlling}. In present, measurement-device-independent quantum key distribution(MDI-QKD)\cite{liu2013experimental,lo2012measurement,ma2012alternative} may be the most promise and attractive QKD protocol, because all potential side channel attacks on measurement device are removed\cite{lo2012measurement}.

In recent years, people did many efforts to further improve the practicality of MDI-QKD. They present decoy state method\cite{wang2013three,yu2015statistical,zhou2016making} and finite-key analysis\cite{curty2014finite,ma2012statistical} to improve the practical security, Reference-Frame-Independent(RFI)\cite{wang2015phase,zhang2017practical} scheme and polarization randomizing scheme\cite{wang2017measurement} to make the MDI-QKD system environmental robustness, etc.

It is quite natural to extend MDI-QKD structure to a "star" network\cite{tang2016measurement,frohlich2013quantum}, in which all users connect to the untrusted relay Charlie. 
The problem is, when calculating MDI-QKD key rate and doing experiments, we always suppose two users have the same distance to Charlie (we could call it 'symmetric situation', and call Charlie's place 'symmetric site'. Otherwise, the 'asymmetric situation' and 'asymmetric site'), which is difficult to reach in the real world, especially in making MDI-QKD network since there are too many users and is nearly impossible to place Charlie to symmetric site.

The secure key rate of MDI-QKD is heavily depend on symmetry level of two users' channel losses since the protocol need Hong-Ou-Mandel(HOM) interference. When losses are asymmetric,the key rate decay fast. Recently, Wenyuan Wang, et al. present a '7-intensity protocol'\cite{wang2018enabling} to deal with the problem. And we will introduce another asymmetric protocol in this paper.

The optimizations of parameters are crucial in QKD. It influences secure key rate significantly\cite{xu2014protocol,wang2018enabling}. In MDI-QKD, each pair of users should keep a set of optimized parameters. Thus in large scale MDI-QKD network, parameter distribution could be a big challenge. For example, there are 10000 users in network, $10000 \times 9999 / 2$, i.e about $5 \times 10^7$ sets of parameters are needed. It may take unbelievably long time to generate them. If we need 1 second(actually may be far more than one second) to optimize a pair of parameters, $5 \times 10^7$ seconds means more than one year! When a new user join the network. We should optimize 10000 pairs of new parameters for him. What's worse, the condition on users' side may change and we should adjust parameters to adapt this changing. This may cost plenty of resources too.

Previously, optimized parameters were generated by 'simulation and iteration\cite{xu2014protocol,wang2018enabling}. This method is mature and accurate ---- if you give enough time and iteration times. However, simulation programs usually cost plenty of time, especially, when running optimization program, you have to iterate simulation program many times. Consider building a large scale MDI-QKD network, when there are a new user link to the network or some users need to re-optimize their parameters for some reasons, we have to optimize too many parameters since each pair of users should have their own optimized parameters.

Another problem brought by large scale MDI-QKD network is reference phase difference($\Delta{\phi}$) and misalignment($e_d$)\cite{tang2014measurement,wang2015phase}
on users' side may change with time. Thus we must recalibrate users' system in time. It's really a big projection with traditional methods.

{Previously, system parameters were re-calibrated by 'stop and scan'}\cite{wang2015phase} {or 'add phase stabilization'}\cite{tang2014measurement}. {The former needs too much time and the latter needs too many additional devices. In this paper, we present a scheme based on Back Propagation Artificial Neural Network(BPNN)}\cite{kohonen1988introduction,rumelhart1986learning} {to solve above problems. BPNN could not only calculate the optimized parameters directly, but also estimate $\Delta{\phi}$ and $e_d$. We could recalibrate the system according to the result given by BPNN rather than stopping and scanning the system or adding phase stabilizations. We validate this idea by simulation and find it is indeed useful.}

The rest of this paper was organized as follows. In Sec.2, we briefly introduce an asymmetric protocol base on 3-intensity protocol\cite{wang2013three,yu2015statistical}, our BPNN training data was generate by this protocol. In Sec.3, we introduce how to make a BPNN to calculate optimized parameters and compare it with traditional local search algorithm (LSA)\cite{xu2014protocol}, and we will discuss the potential of BPNN. In Sec.4, we introduce methods to make and use BPNN to estimate the $\Delta{\phi}$ and $e_d$, we give simulation data to prove the validity of this way. Finally in the appendix A and B, we introduce the working principle of BPNN and process of LSA.

\section{Asymmetric Protocol Based on 3-intensity protocol}

In asymmetric situations, asymmetric protocols perform much better than symmetric version\cite{wang2018enabling,liu2018experimental}. {In this section, we firstly introduce an asymmetric protocol derived from 3-intensity method}\cite{wang2013three,yu2015statistical}. {We will generate our BPNN dataset by this protocol.} Original 3-intensity method considers signal and decoy state intensity in Z-basis($\mu_Z$ and $\nu_Z$) and X-basis($\mu_X$ and $\nu_X$), probability to choose signal and decoy state in Z-basis($P_Z^\mu$, $P_Z^\nu$) and in X-basis ($P_X^\mu$, $P_X^\nu$), i.e. 8 parameters in total. We don't consider probability of vacuum state($P^o$) because $P^o = 1 - P_X^\mu -P_X^\nu - P_Z^\mu -P_Z^\nu$, and don't distinguish vacuum in Z and X-basis since vacuum state don't have any photons.

In our asymmetric protocol, Alice and Bob could choose different parameters. Thus we should optimize 16 parameters totally. We denote these parameters as vector $\vec{v}$ showed in formula(\ref{eq1}). Subscripts a(b) denotes Alice(Bob).

\begin{equation}
\begin{aligned}
\vec{v} = [\mu_{Za},\nu_{Za},\mu_{Xa},\nu_{Xa},P_{Za}^\mu,P_{Za}^\nu,P_{Xa}^\mu,P_{Xa}^\nu,\\
       \mu_{Zb},\nu_{Zb},\mu_{Xb},\nu_{Xb},P_{Zb}^\mu,P_{Zb}^\nu,P_{Xb}^\mu,P_{Xb}^\nu]  
\end{aligned}
\label{eq1}
\end{equation}

 When all system conditions(including channel loss, misalignment, detector efficiency, data size and dark count rate)are fixed, secure key rate($r$) is function of $\vec{v}$, i.e. $ r = R\left( \vec{v} \right)$. Parameters optimization could be regarded as searching for $\vec{v}_{opt}$ to maximize $r$ as showed in Eq.(\ref{eq2}). Where $V$ denotes searching space of $\vec{v}$.

\begin{equation}
\vec{v}_{opt} = arg\ max_{\vec{v}\in{V}}[R(\vec{v})]
\label{eq2}
\end{equation}

\begin{figure}[htbp]

\includegraphics[width=9cm]{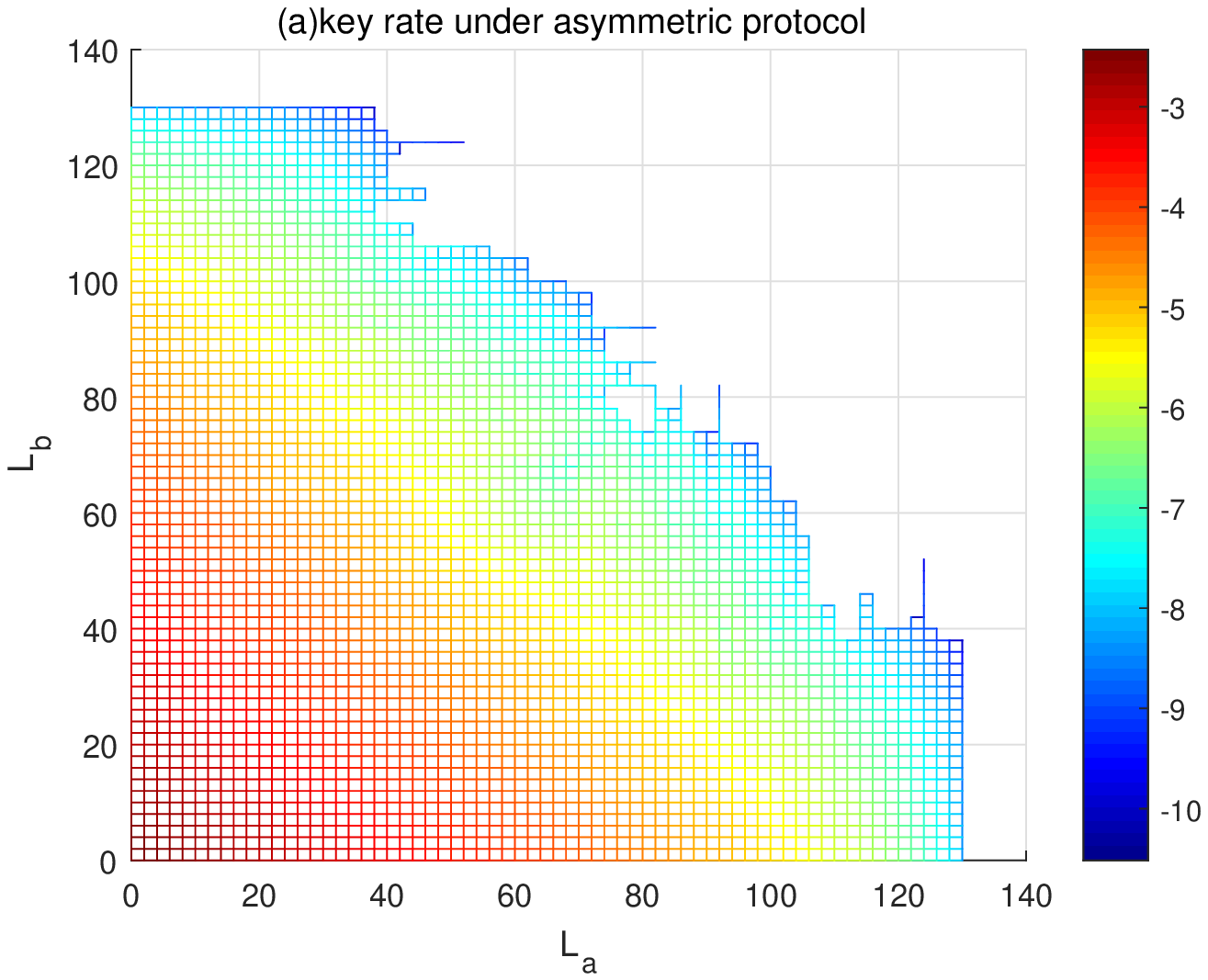}

\caption \\Secure key rate VS users distance of our asymmetric protocol. X(Y) axis denotes distance between Alice(Bob) and Charlie. Color map on the right of mash picture indicates the relation between colors and logarithm of key rate .
\label{fig1.1}
\end{figure}

\begin{figure}[htbp]


\includegraphics[width=9cm]{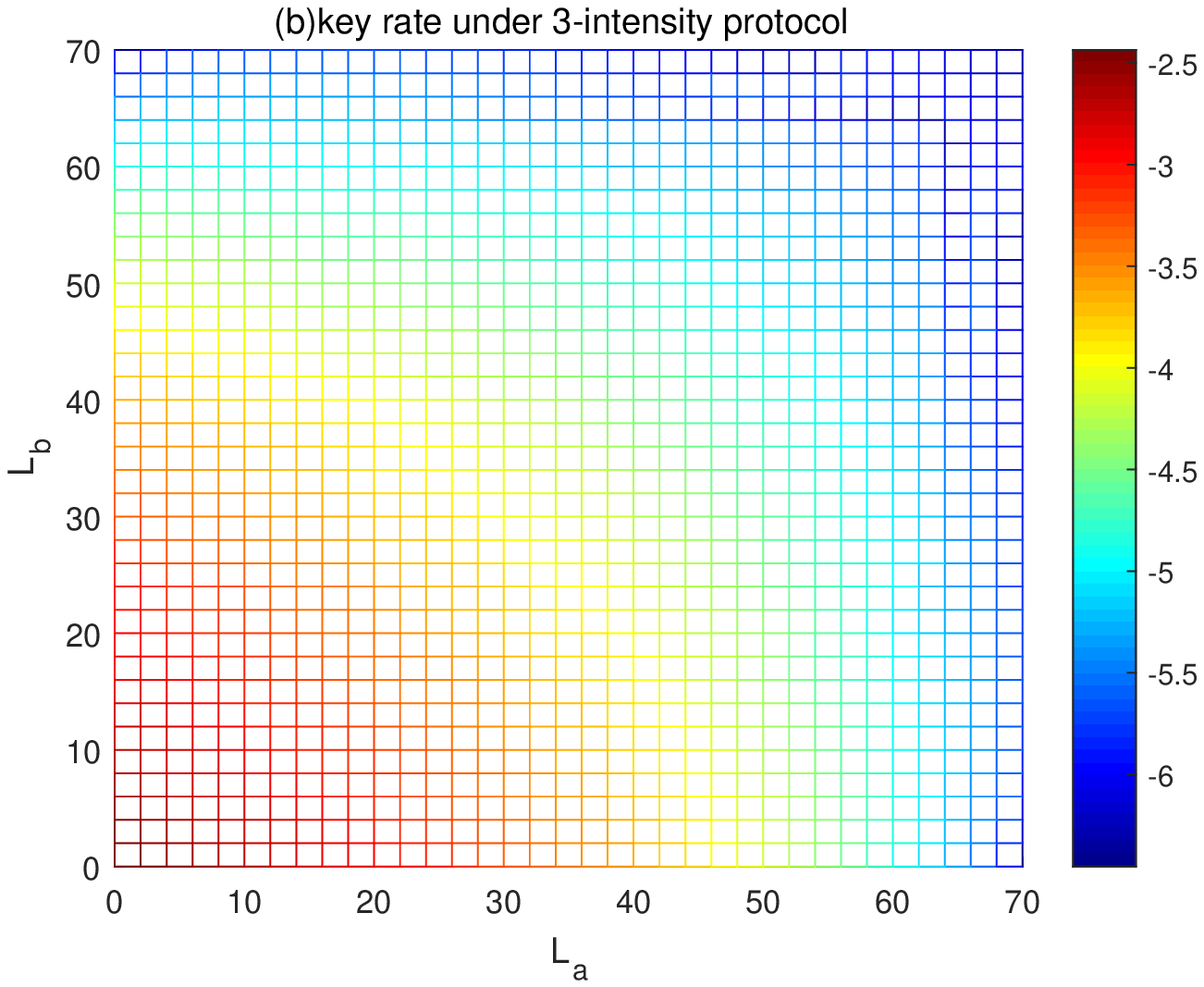}

\caption \\Secure key rate VS users distance of 3-intensity protocol. The meaning of X(Y) axis is same as Fig.\ref{fig1.1}
\label{fig1.2}
\end{figure}

\begin{figure}[htbp]
\includegraphics[width=9cm]{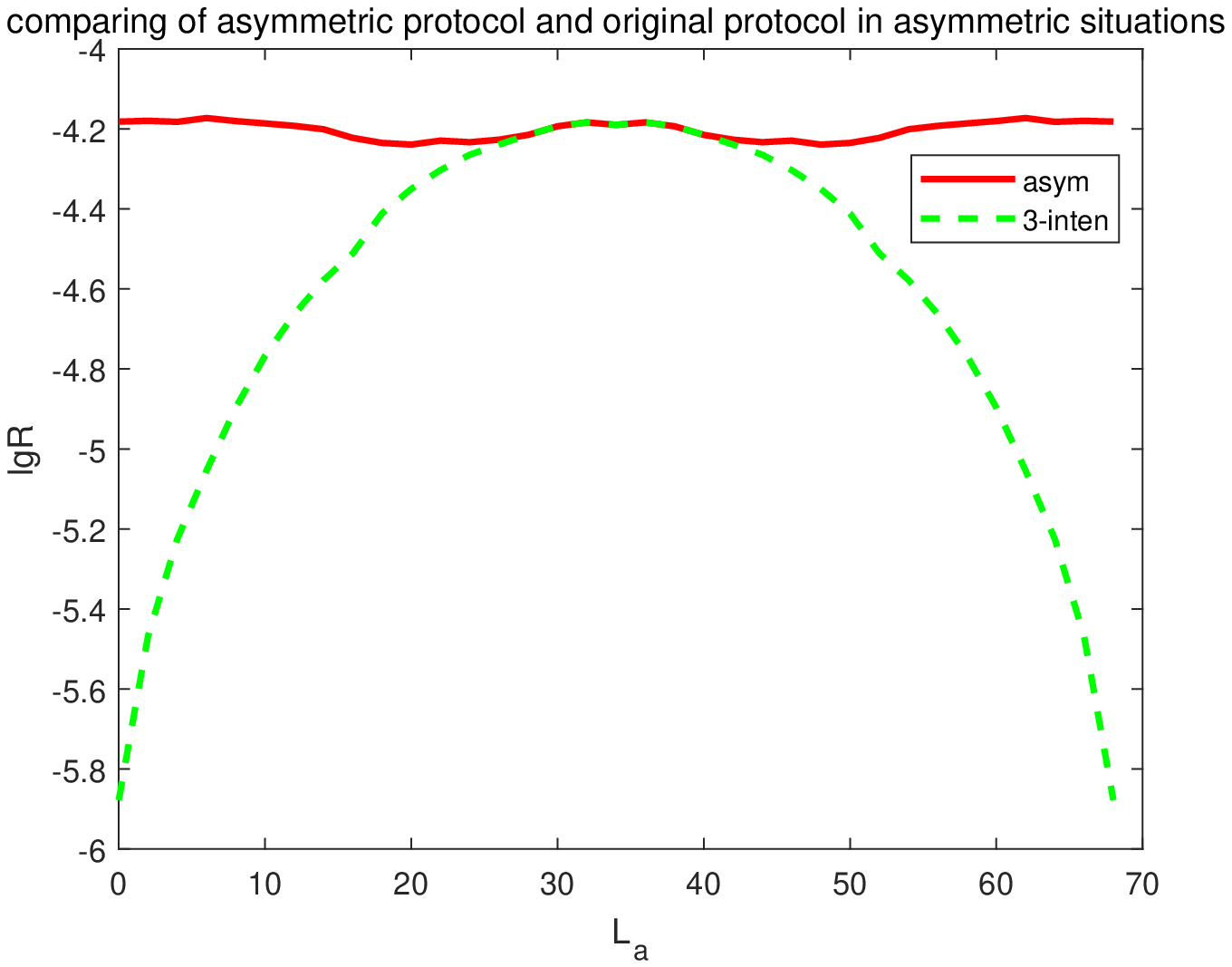}

\caption \\Secure key rate on conditions of $L_a + L_b = 70km$. The red solid line denotes asymmetric protocol and green dash line means 3-intensity protocol.
X axis is distance between Alice and Charlie and Y axis is logarithm of key rate. 
 \label{fig2}
\end{figure}

Here we fix system conditions as showed in Tab.\ref{tab1} and compare key rate($r$) vs channel length($L_a$ , $L_b$) in different protocol. The result is generated from numerical simulation\cite{wang2014simulating} and showed in Fig.\ref{fig1.1}, Fig.\ref{fig1.2} and Fig.\ref{fig2}. Fig.\ref{fig1.1} is the result of our asymmetric protocol, we could find in asymmetric situations, it has higher key rate and reaches a longer distance than 3-intensity method showed in Fig.\ref{fig1.2}. Then we limit $L_a + L_b = 70km$ and compare two protocols. The result is showed in Fig.\ref{fig2} We could find that our protocol performs better than the 3-intensity protocol in strong asymmetric situations.

\begin{table}[htbp]
\centering
\caption{\bf{system conditions for numerical simulation}\\
Where $dc$ denotes dark count, $\eta_d$ denotes detector efficiency, $e_d$ denotes misalignment, $f$ denotes error correcting efficiency and $N$ is data size}
\begin{tabular}{ccccc}
\hline
$dc$ & $\eta_d$ & $e_d$ & $f$ & $N$\\
\hline
$6.02*10^{-8}$ & $70\%$ & $1.5\%$ & $1.16$ & $10^{12}$\\
\hline
 
\end{tabular}
\label{tab1} 
\end{table}

\section{predict optimized parameters by BPNN}
{We use 'predict' in the title to emphasize our method is different from traditional methods. We calculate best parameters directly with the help of machine learning(ML) rather than use simulation and iteration to optimize them. ML is an algorithm that uses statistical techniques to give computer systems the ability to 'learn'. It is widely used in regression, classification, prediction, and clustering, etc.}

Eq.(\ref{eq2}), on the one hand, told us given system conditions, we could find a set of parameters $\vec{v}_{opt}$ to maximize $r$. {It enlighten us there should be  function relationship} between system conditions($\vec{s}$) and $\vec{v}_{opt}$. We express it as Eq.(\ref{eq3}). Where $V$ denotes searching space of $\vec{v}$, $S$ denotes space of $\vec{s}$.

\begin{equation}
\begin{aligned}
\vec{v}_{opt} = f(\vec{s}) ; \vec{v}_{opt}\in{V}, \vec{s}\in{S}
\end{aligned}
\label{eq3}
\end{equation}

Function $f$ in Eq.(\ref{eq3}) is quite difficult to find analytically. But we could approximate its performance by ML according to Universal Approximation Theorem\cite{csaji2001approximation,cybenko1989approximations}, which says simple Artificial Neural Networks(ANN) can represent a wide variety of interesting functions when given sufficient data and appropriate hyperparameters.

\begin{figure}[htbp]
\includegraphics[width=9cm]{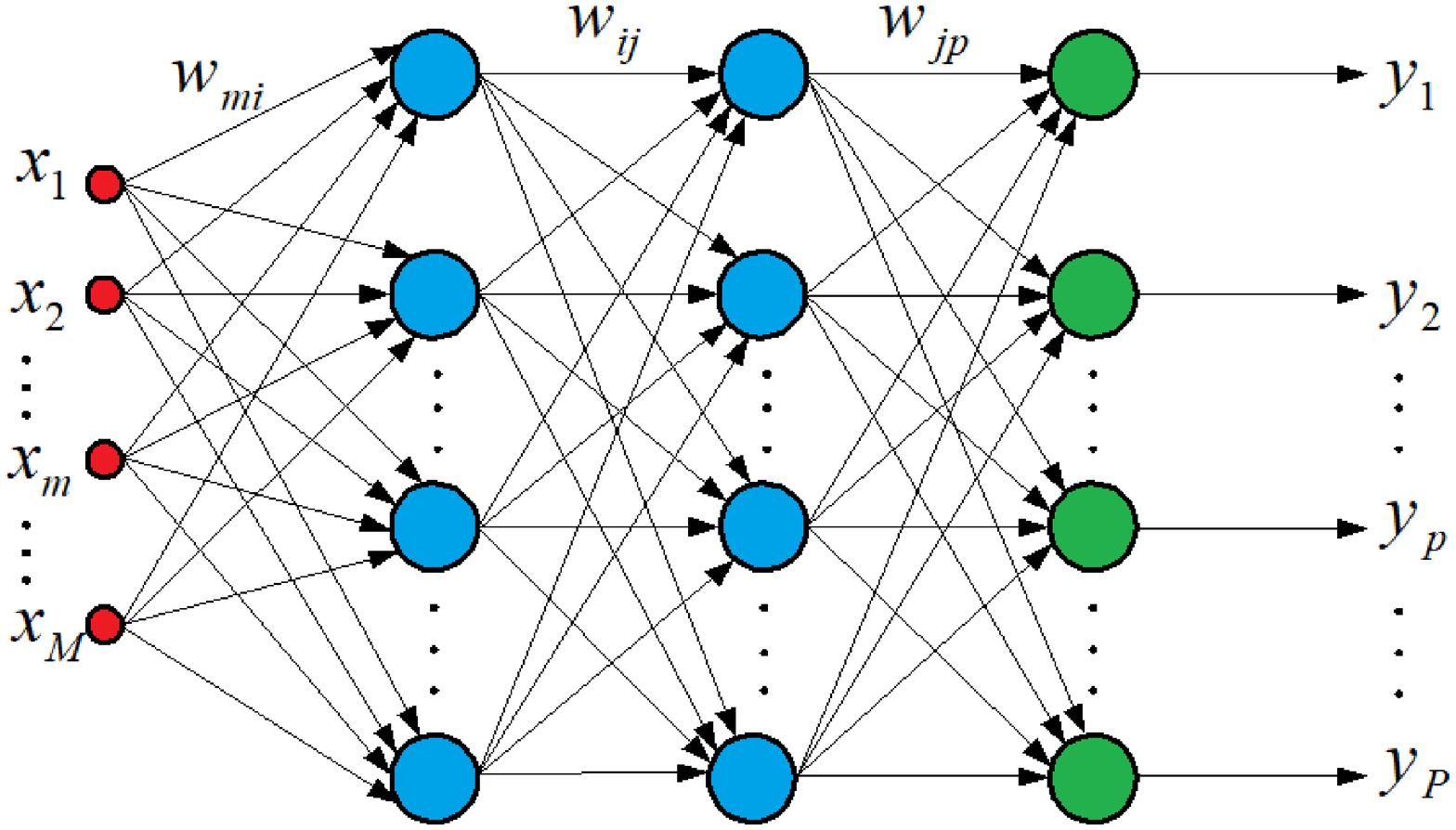}

\caption \\Schematic diagram of BPNN, where red circles denote input layer neurons, blue circle denote hidden layer neurons, green circle denote output layer neurons and black arrow is synapses. $x$ is input parameters, $y$ is output parameters and $w$ are weights of synapses. 
 \label{fig3}
\end{figure}

We choose BPNN, a widely used ANN, to fitting the $f$ since its good ability of nonlinear mapping, self-adapting and generalization. As showed in Fig.\ref{fig3}. BPNN include an input layer, several hidden layers and an output layer. Each layer consists of  
neurons and connects to adjacent layers by synapses.The data recieved by input layer, forward propagate and operated by synapses and nuerals, finally outputted by the output layer.

System conditions $dc$,$\eta_d$,$f$,$N$ are depended on operator's devices. We call them Charlie's conditions and regard them as fixed value. $L_a$,$L_b$ are depended on distance between users and Charlie. $e_d$ is mainly depended on users' misalignment of quantum state preparation. We call them users' conditions. Our purpose is to predict the best parameters according to users' conditions. To improve the precision, each BPNN only predicts one element in $\vec{v}$. Thus we need totally 8 BPNNs, and we will introduce the BPNN which is used to predict $\mu_{Za}$ and $\mu_{Zb}$ as representative in the following.

Firstly we create an input layer with 3 neurons. They receive $L_a$,$L_b$ and $e_d$ as input respectively. Then we add a hidden layer consists of $20 \times 20$ neurons to link with input layer. Finally we place an output layer with only one neurons, which will output the target parameter.

The next step is preparing training data for BPNN. This step uses traditional simulation\cite{wang2014simulating}
{, LSA and Particle Swarm Optimization Algorithm (PSO). The PSO can optimize the non-smooth function and non-convex function, but its speed is very slow. that is to say, PSO can find more accurate optimized parameters. The steps of dataset preparation is introduced as following}

{
\textbf{Step.1} Fixing Charlie's conditions as Tab.}\ref{tab2}.{  Then choose users' conditions as $\vec{u}^0 = (L_a^0, L_b^0, e_d^0)$. Using PSO to find the optimzied point $\vec{v}_{opt}^0$.
}

{
\textbf{Step.2} Changing $\vec{u}^0$ to $\vec{u}^1$, but the difference should be very small. Choose $\vec{v}_{opt}^0$ as initial point and using LSA to optimize the parameters. The result is denoted as $\vec{v}_{opt}^1$. We trust the difference between $\vec{v}_{opt}^0$ and $\vec{v}_{opt}^1$ should be very small since the difference between $\vec{u}^0$ and $\vec{u}^1$ is small. LSA can be faster and more accurate when it has better initial point.
}

{
\textbf{Step.3} repeat Step.2 to generate more pairs of $\vec{u}^i$ and $\vec{v}_{opt}^i$. When $\vec{u}^i$ is special point(such as $L_a = L_a^0$), using PSO algorithm rather than LSA.
}

{
\textbf{Step.4} Record all $\vec{u}^i$ and $\vec{v}_{opt}^i$ as our dataset. 
}

We only use PSO on some special user's conditions. It’s a tradeoff between time and precision.

\begin{table}[htbp]
\centering
\caption{\bf Charlie's conditions}
\begin{tabular}{cccc}
\hline
$dc$ & $\eta_d$ & $f$ & $N$\\
\hline
$6.02 \times 10^{-8}$ & $70\%$ & $1.16$ & $10^{12}$\\
\hline
 
\end{tabular}
\label{tab2} 
\end{table}

The final step is training. We introduce the BPNN which is used to predict $\mu_{Za}$ and $\mu_{Zb}$ as representative. We extract first element ${\mu_{Za}^i}$ from each $\vec{v}_{opt}^i$ and make a label dataset $V_1 = \{\mu_{Za}^i | 1\le{i}\le{M}\}$. Then put dataset $U$ into BPNN, and adjust weight of synapses with the back propagation algorithm(BPA) according to the difference between output of BPNN and label dataset $V_1$. We will introduce the working principle of BPNN in detail in Appendix. It is worth noting that, when predicting $\mu_{Zb}$, we should switch $L_a$ and $L_b$ first. 

\begin{figure}[htbp]
\includegraphics[width=9cm]{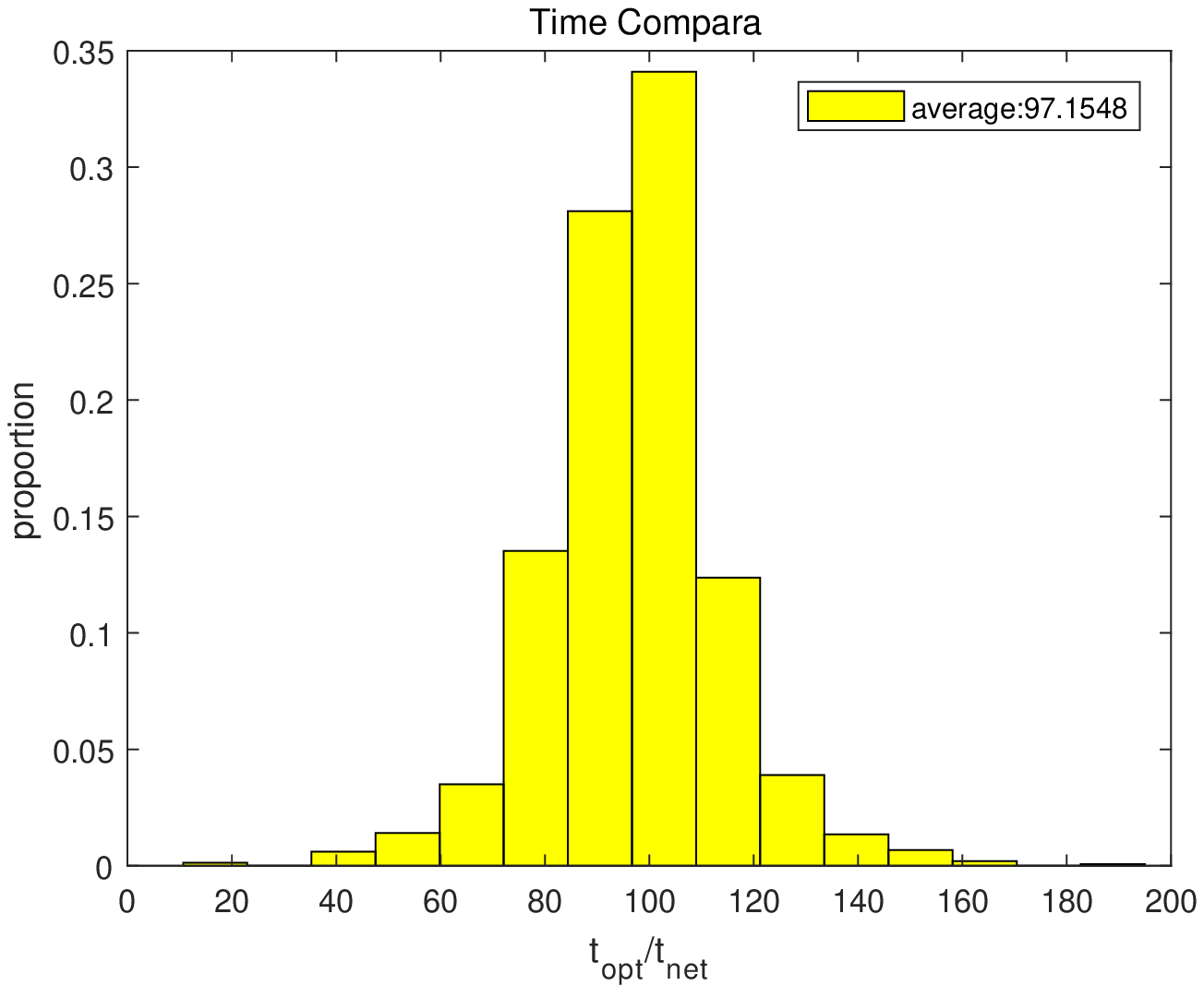}

\includegraphics[width=9cm]{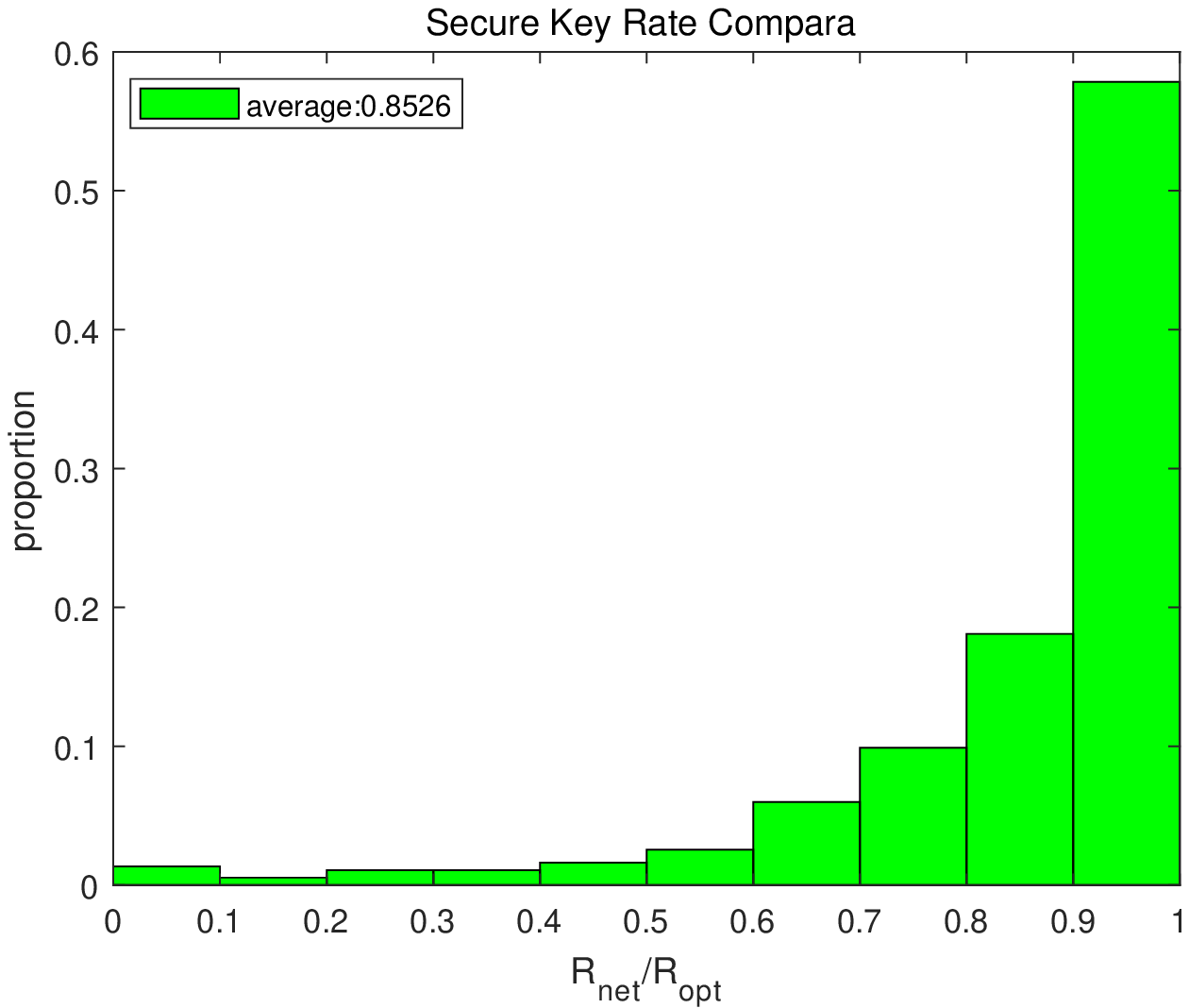}

\caption \\Comparing of running time and secure key rate between two methods. We randomly generate {1500} different user's conditions and calculate best parameters by BPNN and LSA. In the first figure, X axis is running time ratio of LSA to BPNN. Y axis is porperation of this ratio of {1500} data. {The average running time of optimization program and BPNN are, respectively, 17.4653s and and 0.1862s. The average ratio is 97.1548.\\ In the second figure, X axis is key rate ratio of LSA to BPNN, The average ratio is 0.8526.}

 \label{fig4}
\end{figure}

To verify the validity of our BPNN, we randomly generate {1500} sets of users' conditions, then calculate optimized parameters by LSA and BPNN respectively in my PC(CPU:Intel Core i5 7500@ 3.40GHz; RAM: DDR4 8GBytes). 
We compare the running time and key rate.The result is showed in Fig.\ref{fig4}. It indicates that, comparing with troditional LSA, the speed of our method is more than 100 times faster and key rate loss is less than 20\% with the probability of 80\%. The biggest superiority of BPNN is the average time running is only 0.13 second. It's difficult for traditional methods to reach this speed in same hardware condition since the limitation of simulation program.

BPNN has another particular potential that their essence a is simple formula. It means that we could build BPNN by hardware(such as embedded systems) to further improve the speed(specially designed system usually runs faster). But it's nearly impossible to design a hardware system for simulation program since it's too complex.

\section{Real-Time Calibration by BPNN}
In MDI-QKD, reference phase difference between two users $\Delta{\phi}$ lead to errors in X-basis. Quantum states misalignment $e_d$ lead to errors in both Z and X-basis. To make the problem worse, these misalignments may change with time. Thus, MDI-QKD requires us to calibrate users' system in time. As discussed in the introduction, it's quite difficult to accomplish the mission in a large scale network. Consider there are $n$ users in the network. If we try to add phase stabilization, about $n \times (n-1) / 2$ phase stabilizations are needed. If stopping and scanning the system, more than $n$ times scan are needed. Obviously, big $n$ may lead to system unable to work. 

Consider a situation as follows. Two users in the network have fixed distance $L_a$,$L_b$. They optimized their parameters as $\vec{v}_{opt}$ under the conditions of $\Delta{\phi} = 0$ and $e_d = 0.002$. But $\Delta{\phi}$ and $e_d$ increase with time.

With the help of BPNN, we could estimate the $\Delta{\phi}$ and $e_d$ in real time, only need to statistic data of single photon error rate in X-basis ($e_{11}^X$), qubit error rate of signal state in Z-basis ($E_{\mu}^Z$) and secure key rate ($r$), instead of adding additional devices or stopping the system. It's worth noting that these data are calculated for error correction and privacy amplification originally, i.e. they are recycled, not generate for BPNN specially.

An other superiority of our method is speed. BPNN could judge how much $\Delta{\phi}$ and $e_d$ are there in users' system accurately and quickly. If we have known the value, we could recalibrate users' system immediately.

{Apart form these, there are two main superiorities compared with using traditional curve fitting to estimate $\Delta{\phi}$ and $e_d$. The first is the ability 'Non-linear mapping'. there is a theory called Universal approximation theorem}
\cite{csaji2001approximation,cybenko1989approximations}.
{It states that a feed-forward network with a single hidden layer containing a finite number of neurons can approximate continuous functions on compact subsets, under mild assumptions on the activation function. The theorem thus states that simple neural networks can represent a wide variety of interesting functions when given appropriate parameters. When using traditional curve fitting, we should have some priori knowledge about the data we want to fitting and carefully choose a mode such as linear, logarithmic or polynomial etc. But, when using BPNN, we don't care about the mode, we just prepare our dataset and push them to BPNN to do the training.}

{Another one is the ability to process high dimensional data. Artificial neural networks can easily process high dimensional data such as a photo with millions dimension. But curve fitting can't be up to this job. Although in the problem of estimating $\phi$ and $e_d$, the dimension is not too high (input is three dimension and output is two dimension). But we want to give a demonstration and inspiration to others. Maybe in the future, we can use more input data to correct the polarization misalignment, time misalignment etc. And people may find more regular of QKD system by AI algorithm.}

The BPNN for calibration need $e_{11}^X$, $E_{\mu}^Z$, and $r$ as its input. $e_{11}^X$ mainly implies the $\Delta{\phi}$, $E_{\mu}^Z$ implies the $e_d$, and $r$ is sensitive to all kinds of errors. Thus we choose them to be our input and use BPNN to find their relations.

Firstly we prepare our training dataset. The process is showed as following:

{\textbf{Step.1} We fix Charlie's conditions and $L_a$ $L_b$ as showed in Tab.}\ref{tab3}.{ Choosing phase misalignment and misalignment error as $\Delta{\phi}^0$ and $e_d^0$. Then we simulate the MDI protocol, get the $e_{11}^{X0}, E_{\mu}^{Z0}, r^0$}

{
\textbf{Step.2} Changing phase misalignment and misalignment error to $\Delta{\phi}^1$ and $e_d^1$. Then we simulate the MDI protocol, get the $e_{11}^{X1}, E_{\mu}^{Z1}, r^1$
}

{
\textbf{Step.3} Repeating the step.2 and record all $\Delta{\phi}^i$, $e_d^i$, $e_{11}^{Xi}, E_{\mu}^{Zi}, r^i$ as our training dataset. The $\Delta{\phi}^i$, $e_d^i$ are our output dataset and $e_{11}^{Xi}, E_{\mu}^{Zi}, r^i$ are our input dataset.
}

\begin{figure}[htbp]
\includegraphics[width=9cm]{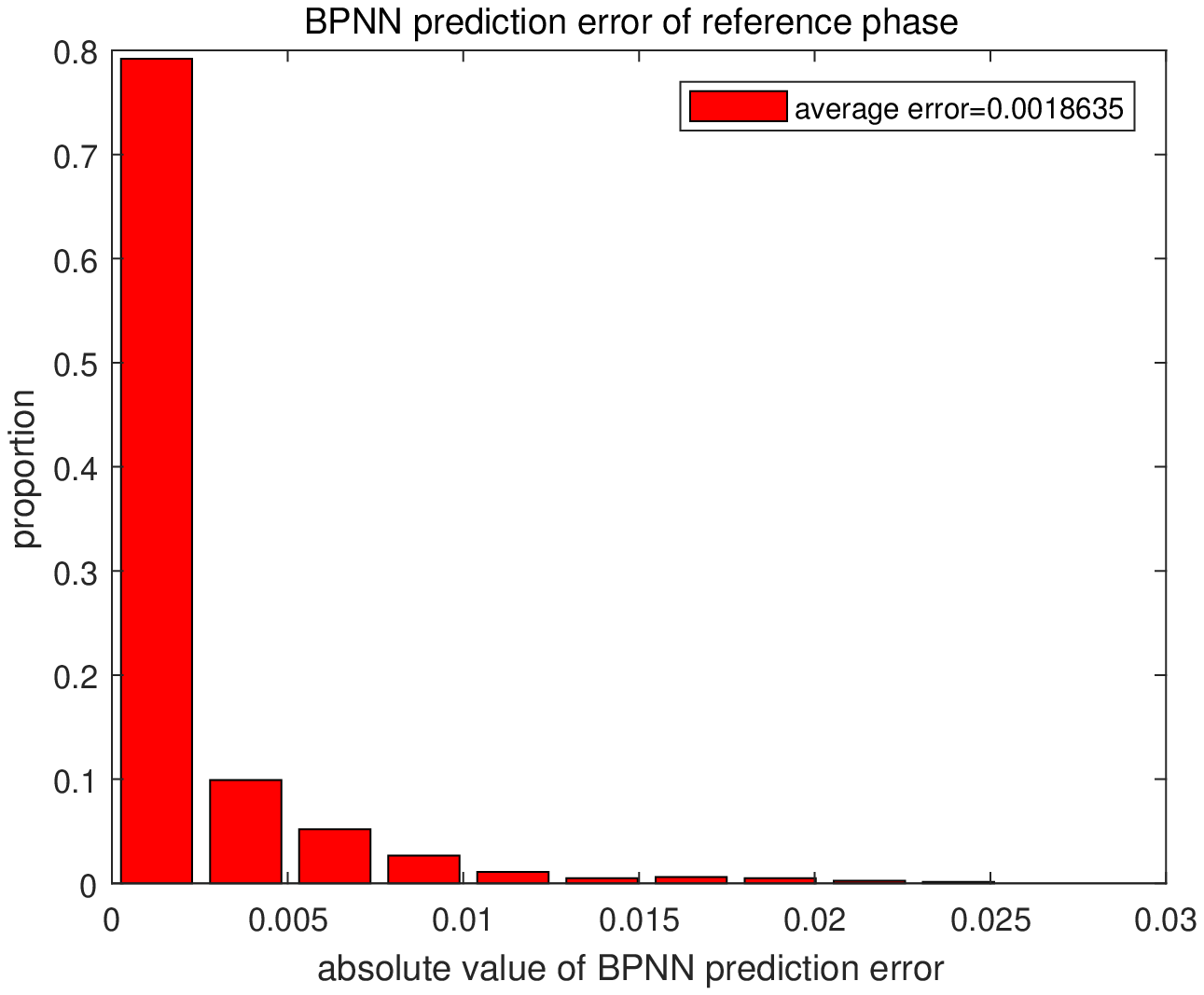}

\includegraphics[width=9cm]{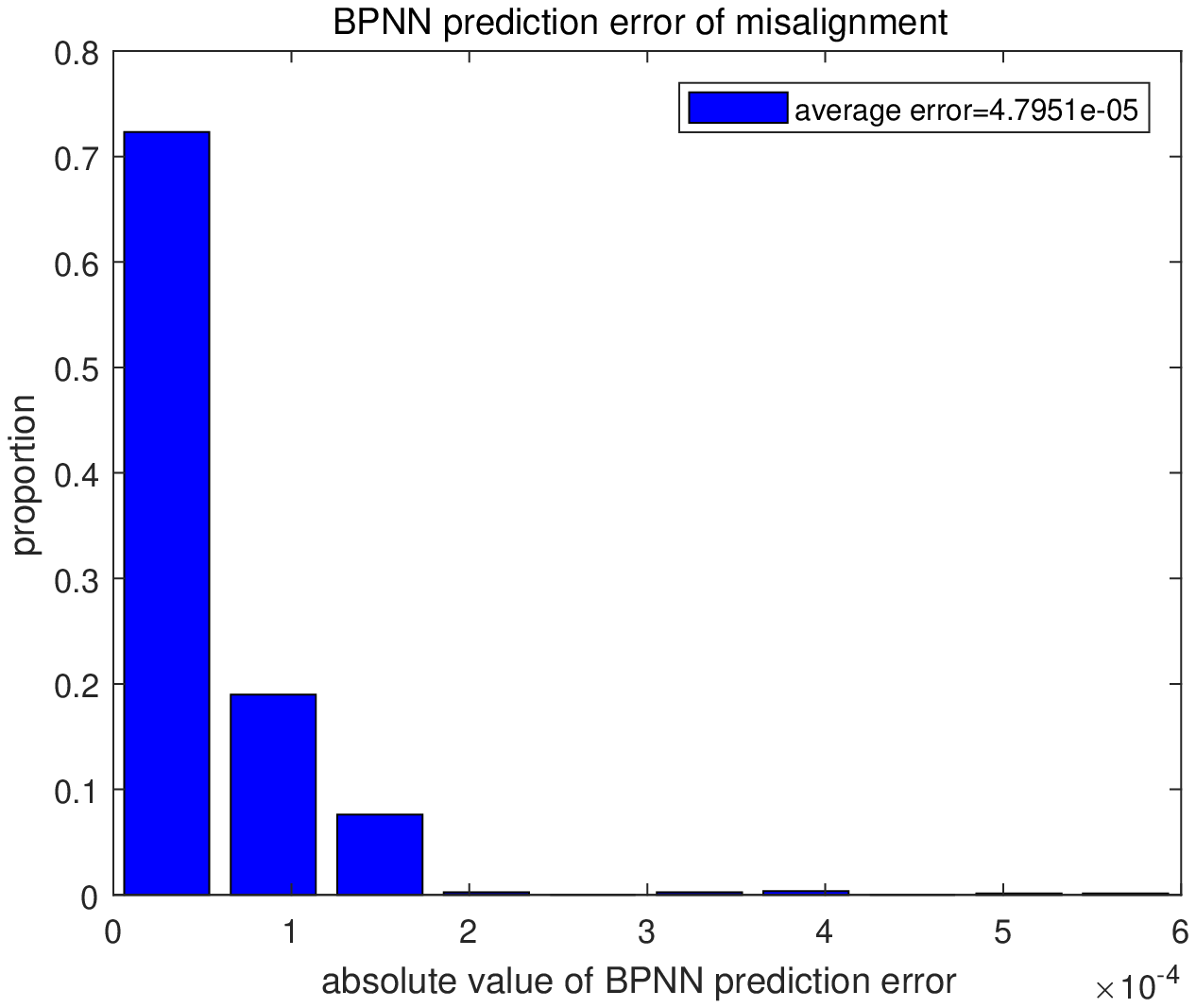}

\caption \\The BPNN prediction error of $\Delta{\phi}$ and $e_d$. In first figure, X axis is the absolute value of the difference between $\Delta{\phi}$ and $\Delta{\phi}_{net}$. Y axis is the proportion of data. The average error is only 0.0018 while the range of $\Delta{\phi}$ is 0 to 0.5.\\ 
In second figure, X axis is absolute value of difference between $e_d$ and $e_{d\_net}$. Y axis is the proportion of data, the average error is $4.79\times10^{-5}$ while the range of $e_d$ is 0.002 to 0.02.

 \label{fig5}
\end{figure}

{To verify the validity, we randomly generate 1000 pairs of $\Delta{\phi}^i$ and $e_d^i$, where $\Delta{\phi}$ is uniformly distribute in 0 to 0.5 and $e_d$ is uniformly distribute in 0.002 to 0.02. Then we run the simulation program to obtain related $e_{11}^{Xi}$, $E_{\mu}^{Zi}$ and $r^i$. We put $e_{11}^{X}$, $E_{\mu}^{Z}$ and $r$ into BPNN, then BPNN estimate these misaligments, denoted by $\Delta{\phi}_net$ and $e_{d\_net}$. We compare the difference between true misaligments and estimated misaligments.} The result is showed in Fig.\ref{fig5}. It indicate that difference between estimated value and the true value is about 2 order of magnitude smaller than the true value, i.e. the estimation is quite accurate.

\begin{table}[htbp]
\centering
\caption{\bf system conditions}
\begin{tabular}{cccccc}
\hline
$L_a$ & $L_b$ & $dc$ & $\eta_d$ & $f$ & $N$\\
\hline
$10km$ & $20km$ & $6.02 \times 10^{-8}$ & $70\%$ & $1.16$ & $10^{12}$\\
\hline
\end{tabular}
\label{tab3} 
\end{table}

\section{Discussion}
{In summary, We firstly introduce a method to develop symmetric 3-intensity protocol to adapt the asymmetric situations and prove it indeed has higher key rate and reaches a longer distance by simulation. Our BPNN training dataset is generated by this method. }

Then, We presented a totally new idea to get optimized parameters. With the help of BPNN, we don't {need to} use optimization algorithm any more. We verified the performance of BPNN and found it could run far faster than traditional methods and analyzed the reason for its faster speed in principle. We finally discussed the potential of BPNN which optimization algorithm could never reach.

Also, we found BPNN could support to calibrate the system in time, and verified it by simulation. In simulation results, the estimations of BPNN are very accuracte. We trust this is a valuable idea since at present, calibration methods cost too much additional time or devices. Our method gives a possibility to improve the system performance only with discarded data.

\section{ACKNOWLEDGMENTS} 
This work has been supported by the National Key Research and Development Program of China (2016YFA0302600), the National Natural Science Foundation of China (NSFC) (61775207, 61575183, 61771439, 61702469); Strategic Priority Research Program (B) of the Chinese Academy of Sciences (CAS) (XDB01030100, XDB01030300), and the Foundation of Science and Technology on Communication Security Laboratory (6142103040105).

\section{Appendix A : Working principle of Back Propagation Artificial Neural Network}

Artificial neural networks (ANN) consist of simple units with adaptability connect with each other. It is inspired by biological neural networks. The simple units come from neurons and connections come from synapses. A neuron could receive a signal, process it and transmit it to additional neurons connected to it. Error Backpropagation Algorithm\cite{rumelhart1986learning} is used to calculate a gradient that is needed when training weights of synapses. The combination of them are called Back Propagation Artificial Neural Network(BPNN).

We will introduce BPNN by Fig.\ref{fig3}. The red circles are input neurons, they receive $x_m$ as input signals, where $m = 1,2,....,M$. Then the signals go through the synapses with certain weights and summed by hidden neurons' input side as showed in Eq.(\ref{eq4}). 

\begin{equation}
  \epsilon_i = \sum_{m=1}^{M}{w_{mi}\times{x_m}}
\label{eq4}
\end{equation}

Then, the $i$'th hidden neurons, denoted by blue circles, receive $epsilon_i$ as its input signals and process them by active function $f$ as showed in Eq.(\ref{eq5}). Where $\theta$ is threshold.

\begin{equation}
  \zeta_i = f(\epsilon_i - \theta_i)
  \label{eq5}
\end{equation}

The following steps are the same as above, $zeta_i$ go through the synapses with certain weights and summed as inputs of the next layer, processed buy active functions and so on, until the output layer gives results $y$ as Eq.(\ref{eq6}).

\begin{equation}
  \epsilon_p = f(\sum_{j=1}^{P}{w_{jp}\times{\zeta_j}}-\theta_p)
  \label{eq6}
\end{equation}

\section{Appendix B : Using Local Search algorithm to optimize parameters}

We use LSA to generate optimized parameters as training data of BPNN. The process is described as follows.

Step.1 Setting up precision value $\tau$ to shut down the algorithm and step length $s$ to search better value.

Step.2 We choose a good initial point $\vec{v}$ as showed Eq.(\ref{eq1}) and calculate the key rate by simulation program, i.e. $r = S(\vec{v})$ , where $S$ is simulation program which receive parameters $\vec{v}$ as input and output key rate $r$.

Step.3 The first element in $\vec{v}$ is $\mu_{Za}$, we move it to $\mu_{Za} + \Delta{s}$ and $\mu_{Za} - \Delta{s}$. Then we get two new points $\vec{v}_{+}$ and $\vec{v}_{-}$ as Eq.(\ref{eq7}) and calculate key rate at these new points as Eq.(\ref{eq8}).

\begin{equation}
\begin{aligned}
\vec{v}_{\pm} = [\mu_{Za} \pm \Delta{s},\nu_{Za},\mu_{Xa},\nu_{Xa},P_{Za}^\mu,P_{Za}^\nu,P_{Xa}^\mu,P_{Xa}^\nu,\\
       \mu_{Zb},\nu_{Zb},\mu_{Xb},\nu_{Xb},P_{Zb}^\mu,P_{Zb}^\nu,P_{Xb}^\mu,P_{Xb}^\nu]  
\end{aligned}
\label{eq7}
\end{equation}

\begin{equation}
r_{\pm} = S(\vec{v}_{\pm})
\label{eq8}
\end{equation}

Step.4 Compare $r$, $r_{+}$ and $r_{-}$. if $r_{+}$($r_{-}$) is the maximum one, move our searching point to $\vec{v}_{+}$($\vec{v}_{-}$), i.e. $\vec{v} = \vec{v}_{+}(\vec{v}_{-})$. Else if $r$ is the maximum, keep $\vec{v}$ unchanged. And record the maximum key rate as $r_{max}$.

Step.5 We compare $r$ and $r_{max}$, if the difference of them is less than $\tau$, shut down the algorithm and output $\vec{v}$ as optimized parameters. Else, let $r = r_{max}$ and repeat step.3 to step.5, but move the next element in $\vec{v}$ this time. If it has been the last element in this loop, return to the first element to move.

It is worth noting that, we should choose $s$ and $\tau$ carefully to balance the precision and speed.


\appendix

\nocite{*}

\bibliography{apssamp}

\end{document}